\documentclass[%
 reprint,
superscriptaddress,
 amsmath,amssymb,
 aps,
]{revtex4-2}

\usepackage[pdftex,colorlinks=true,linkcolor=blue,bookmarks=false,citecolor=blue,urlcolor=blue,hypertexnames=false]{hyperref}
\usepackage{graphicx}% Include figure files
\usepackage{dcolumn}% Align table columns on decimal point
\usepackage{bm}% bold math
\usepackage{siunitx} % for nice unit handling
\usepackage{float}
\usepackage{xcolor}
\usepackage{hyperref}

\begin{document}

\title{Reversible phasonic control of a quantum phase transition in a quasicrystal}

\author{Toshihiko Shimasaki}
\author{Yifei Bai}
%\thanks{Equal contribution.}
\author{H. Esat Kondakci}
%\thanks{Equal contribution.}
\author{Peter Dotti}
\author{Jared E. Pagett}
\author{Anna R. Dardia}
\author{Max Prichard}
\affiliation{Department of Physics, University of California, Santa Barbara, California 93106, USA}
\author{Andr\'e Eckardt}
\affiliation{Technische Universität Berlin, Institut für Theoretische Physik, Hardenbergstraße 36, Berlin 10623, Germany}
\author{David M. Weld}
\email{Email: weld@ucsb.edu.}
\affiliation{Department of Physics, University of California, Santa Barbara, California 93106, USA}
 
\begin{abstract}
%I think we should begin with something like: 
Periodic driving can tune the quasistatic properties of quantum matter.
A well-known example is the dynamical modification of tunneling by an oscillating electric field. Here we show experimentally that driving the phasonic degree of freedom of a cold-atom quasicrystal can continuously tune the effective quasi-disorder strength, reversibly toggling a localization-delocalization quantum phase transition. Measurements agree with fit-parameter-free theoretical predictions, and illuminate a fundamental connection between Aubry-Andr\'e localization in one dimension and dynamic localization in the associated two-dimensional Harper-Hofstadter model.
These results open up new experimental possibilities for dynamical coherent control of quantum phase transitions. 
\end{abstract}

\maketitle

Driving can modify the properties of quantum matter~\cite{RevModPhys.89.011004}, tune tunneling~\cite{arimondo-shakenlattice,oberthaler-drivendoublewells}, and control both dynamic~\cite{dunlap-dynloctheory,holthaus-dynloctheory_DL} and Mott localization~\cite{arimondo-coherentcontrol}. While such phenomena have mostly been explored in the context of periodic crystals, richer possibilities exist in non-translationally-symmetric matter. Quasicrystals, which lack both translation symmetry and true disorder, support ``phasonic'' modes not present in ordinary crystals~\cite{shankari-phasonicspectroscopy,kraus_quasiperiodicity_2016, zilberberg_topology_2021, fan_topological_2022} as a consequence of their intrinsic connection to a higher-dimensional superspace~\cite{harper_single_1955, kraus_topological_2012_pumping, kraus_quasiperiodicity_2016}, and can exhibit an Anderson-like Aubry-Andr\'e localization phase transition driven by quasi-disorder~\cite{aubry1980analyticity,roati_anderson_2008_AAExp}. These properties open up fundamentally new possibilities in the exploration of driven matter.

In this work we demonstrate experimentally and confirm theoretically that driving a phasonic degree of freedom in a cold-atom quasicrystal can tune the effective quasidisorder strength and reversibly control a localization quantum phase transition.  As we show, this can be viewed as phasonic Floquet engineering of Aubry-Andr\'e localization in a 1D quasicrystal, or, equivalently, as tunable dynamic localization by an oscillating electric field in the higher-dimensional quantum Hall system from which the quasicrystal is mapped. These results and complementary perspectives illuminate fundamental connections between apparently different forms of localization, and open up new possibilities for Floquet-engineered matter and dynamical quantum simulation.

\begin{figure}[b!]
	\centering
 	\includegraphics[scale=1]{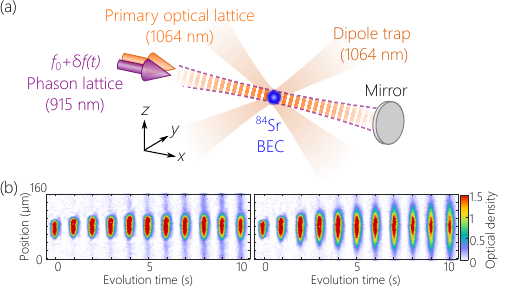}
	\caption{Experimental schematic and typical data. \textbf{(a)} An optically trapped BEC is loaded into a bichromatic lattice and allowed to evolve. A time-varying phasonic displacement between the two sublattices is controlled by varying the frequency of the secondary lattice laser ($\lambda_S = 915~\mathrm{nm}$). 
\textbf{(b)} Absorption images of the atoms taken after various evolution times in the phasonically modulated bichromatic lattice, in the localized regime (left panel) and delocalized regime (right panel).}
	\label{Fig1}
\end{figure}

The experiments we describe begin by loading an optically-trapped Bose-Einstein condensate of $\approx$200,000 $^{84}$Sr atoms into a bichromatic optical lattice composed of a primary lattice with wavelength $\lambda_\mathrm{P}=1063.9774(23)$~nm and a secondary lattice with variable depth $V_S$ and wavelength $\lambda_\mathrm{S}=914.4488(17)$~nm (Fig.~1). Ultracold atoms in specialized optical lattices such as this have been shown to provide an ideal platform for the study of quasicrystals \cite{kevinQCpaper,shankari-phasonicspectroscopy,SchneiderQuasicrystalPRL1,SchneiderQuasicrystalPRL2}. The experiment is initiated by suddenly extinguishing the confining optical dipole trap after ramping up the bichromatic lattice. This realizes the tight-binding Aubry-Andr\'e-Harper (AAH) Hamiltonian
\begin{equation}
	\hat{H}\! =\! -J \sum_{i=1}^L \! \left( \hat{b}^{\dag}_{i} \hat{b}_{i + 1} \! + \! \mathrm{h.c.} \right) + \Delta \! \sum_{i=1}^L \cos\left[2 \pi \alpha i + \varphi(t)\right] \hat{b}^{\dag}_{i} \hat{b}_{i}, \label{AAHamiltonian}
\end{equation}
where $J$ is the tunneling energy which gives rise to a tunneling time $T_J\!=\!\hbar/J$, $\hat{b}^{\dag}_{i}$($\hat{b}_{i}$) is the bosonic creation (annihilation) operator at the $i$-th lattice site, $\Delta$ is the secondary lattice depth, $\alpha=\lambda_\mathrm{P}/\lambda_\mathrm{S}$ is the wavelength ratio of the two lattices, and $\varphi(t)$ is the potentially time-dependent relative phase between the two lattices. This phasonic degree of freedom is controlled by modulating the secondary lattice laser and measured by an interferometer. 
For $\varphi(t)=0$ or almost any constant~\cite{jyotirmskaya}, 
%For  $\varphi(t)=0$ or other constant phases, 
this Hamiltonian exhibits a quantum phase transition at $\Delta=2J$ between localized and delocalized phases \cite{aubry1980analyticity}. When driving the system, in order to avoid the strong interband excitation observed for phasonic driving in \cite{shankari-phasonicspectroscopy}, we choose modulation frequencies in the optimal frequency window \cite{eckardt-optimalfreq} where the modulation is fast compared to the band width but slow compared to the band gap. 

%\section{RESULTS}
% Figure Info
% Resonances 
% 5/20 data: 942 Hz >> Lattice depths 10Er, 0.5Er, hold time 10s, 5 averaging 
% 5/23 data: 314 and 628 Hz >> same parameters 
% Bessel 
% 6/23 data: 314 Hz, Lattice depths 8.5Er and 0.124Er, hold time 9 seconds
\begin{figure}[!htb]
	\centering
	\includegraphics[scale=1]{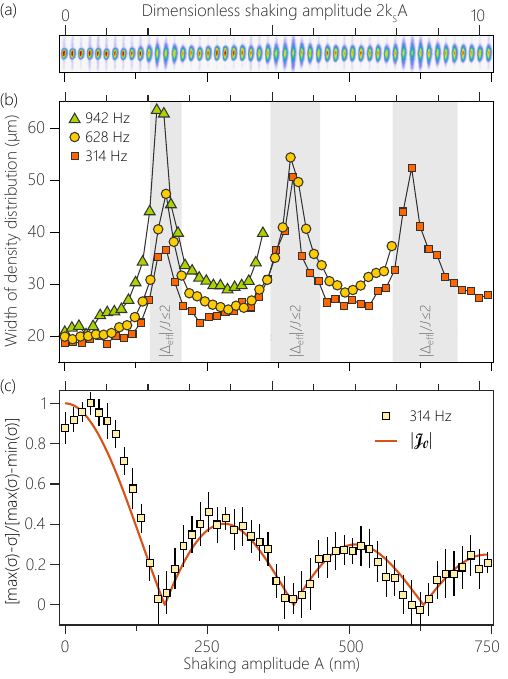} 
	\caption{Phasonic modulation causes dynamic delocalization. \textbf{(a)} Absorption images of the atomic density distribution after 10 s evolution for varying amplitudes of a $314~\mathrm{Hz}$ phason modulation, showing peaks in the late-time width at several drive amplitudes.  \textbf{(b)} Width of the atomic density distribution after 10 s evolution versus phason modulation amplitude, for three different driving frequencies. The delocalized regions are observed to be independent of drive frequency. The primary and secondary lattice depths are 10 Er and 0.5 Er. %, respectively. 
 Here $E_\mathrm{r}=h^2/2m\lambda_\mathrm{P}^2$ is the recoil energy, $m$ is the atomic mass, and $h$ is Planck's constant. 
Shaded areas show the regime of theoretically predicted delocalization described in the text. \textbf{(c)} Quasi-disorder strength can be inferred from transport. Plot shows a normalized form of the late-time width $\sigma$ versus phason modulation amplitude, for primary (secondary) lattice depth $8.5$ and ($0.124$)$~E_\mathrm{r}$, corresponding to the delocalized regime. In this regime the expansion speed is approximately proportional to the quasidisorder strength, so the expected functional form is the absolute value of a Bessel function $|J_0(k_SA)|$, shown here as a solid line with no fit parameters.
All of the panels share the same $x$-axis scaling, measured in the dimensionless shaking amplitude $2k_SA$ (panel top) and in the actual shaking amplitude $A$ (panel bottom).
 	}\label{Fig2}
\end{figure}

We investigate transport in the phasonically driven AAH model by imaging the width $\sigma$ of the atomic density distribution after some evolution time using in-situ absorption imaging. Here $\varphi(t)= 2k_sA\sin(\omega t)$, where $k_S=2\pi/\lambda_S$, $A$ is the phason modulation amplitude, and $\omega$ is the phason
modulation frequency. A natural question to explore is what happens as the amplitude of phasonic modulation is increased from zero in a regime where the unmodulated system is localized. The first main experimental result of this work is shown in Figure~2: as the amplitude of phason modulation is increased, the late-time width is greatly enhanced, indicating delocalization, but only at certain modulation amplitudes. The system appears to switch back and forth between localized and delocalized phases as the drive amplitude increases, with late-time width a non-monotonic function of phason drive amplitude. A crucial clue to the origin of these delocalization peaks is that their positions are independent of drive frequency, in clear contrast both to dynamic localization~\cite{arimondo-shakenlattice} and to expected heating behaviors.

To understand this somewhat counter-intuitive result, it is helpful to expand the second term in Eq.~\ref{AAHamiltonian} \cite{shankari-phasonicspectroscopy}:
\begin{multline}
\Delta \cos(2\pi\beta i + 2k_sA\sin(\omega t)) \\
% \Delta \cos(2k_s [x-A\sin(\omega t)]) \\
%=\sum_{n=-\infty}^\infty J_n(2k_sA) V_s[\cos(2 k_s x)\cos(n\omega t)+\sin(2 k_s x)\sin(n\omega t)]\\
% =\sum_{n=-\infty}^\infty \Delta J_n(2k_sA) \cos(2 k_sx-n\omega t),
=\Delta \sum_{n=-\infty}^\infty J_n(2k_sA) \cos(2 \pi \beta i-n\omega t),
\end{multline}
where $J_n$ are Bessel functions of the first kind.
Keeping only the static $n=0$ term gives rise to a modified effective pseudo-disorder strength $\Delta_\mathrm{eff}$ given by the product of $\Delta$ and $J_0(2k_SA)$. The primary effect of phason modulation is thus to renormalize the effective strength of the incommensurate potential, which becomes a non-monotonic function of the drive amplitude. If the effective quasi-disorder strength falls below $2J$, the system undergoes a quantum phase transition of the Aubry-Andr\'e type into the delocalized phase.  Figure 2 provides support for this interpretation of the results: the gray shaded areas in Figure 2a indicate the predicted regions of phason drive amplitude where $\Delta_\mathrm{eff}<2J$, and correspond very well to the observed delocalization peaks.

As a more quantitative probe of the Floquet-induced rescaling of quasi-disorder, we measured transport starting in a delocalized regime with a lower secondary lattice depth. In this regime, the speed of the ballistic expansion is approximately proportional to the dimensionless distance from the localization phase transition $2-\Delta_{\mathrm{eff}}/J$, a feature we have confirmed numerically~\footnote{See supplementary information.}. For this reason, a plot of the appropriately normalized late-time width of the density distribution as a function of $k_SA$ should take on the exact form of a Bessel function, with an absolute value since such transport measurements do not distinguish positive from negative quasi-disorder. Figure 2b shows just such a plot of normalized late-time widths; the measured data are overall in excellent agreement with a $|J_0|$ Bessel function without any fit parameters. The slight theory-experiment disagreement at low shaking amplitudes is not completely understood.

% Figure Info
% 6/23 data, Lattice depths 6Er, 0.5Er, 314 Hz
% 5 averages
\begin{figure*}[ht]
	\centering
	\includegraphics[width=\textwidth]{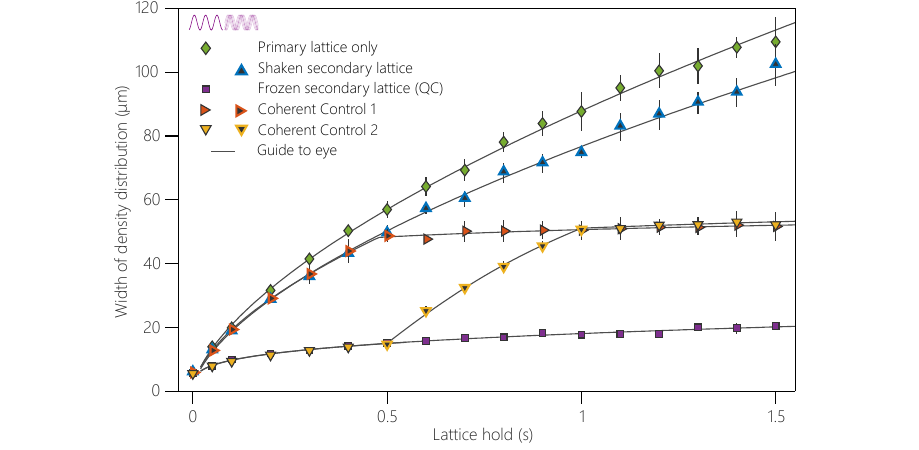} 
	\caption{Reversible coherent control of localization. Symbols show measured late-time width of the density distribution versus hold time for five different experimental protocols: no secondary lattice (diamonds), continuous phasonic driving of secondary lattice (upward triangles), phasonic driving for the first 500 ms (rightward triangles), phasonic driving only between 500 and 1000 ms (downward triangles), and no driving of the secondary lattice (squares).
 For all protocols the primary lattice depth is  $6~E_\mathrm{r}$, and for all but the first plot the secondary lattice depth is $0.5~E_\mathrm{r}$.
 At these values in the absence of driving the system is Aubry-Andr\'e localized. Note especially that width evolution under the second ``coherent control''  protocol shows evidence of localization for times less than 0.5~s and greater than 1~s, and evidence of delocalization between those times, indicating reversible coherent control. 
Shaking frequency is $628~\mathrm{Hz}$ and phason amplitude is $2k_S A \approx 5.52$, near the second Bessel zero. }
	\label{Fig3}
\end{figure*}

An intriguing connection emerges when these results are interpreted in terms of the higher-dimensional superspace associated with any quasiperiodic system. The 1D AAH model can be obtained by dimensional reduction from the 2D anisotropic Harper-Hofstadter model describing a 2D electron gas in a high magnetic field~\cite{harper_single_1955, hofstadter_energy_1976_HH}, in a gauge where the vector potential $\mathbf{A} = (0, 2\pi\alpha i + \varphi(t), 0)$ and zero scalar potential. Here the quasi-disorder strength $\Delta$ becomes the tunneling strength along the extra dimension in the superspace, the incommensurate ratio $\alpha$ describes the magnetic flux per plaquette, and the time derivative $-\partial_t\varphi (t)$ of the phasonic parameter appears as an applied electric field along the extra dimension. The sinusoidal modulation $\varphi(t) = 2k_sA \sin \omega t$ thus corresponds to a driven Harper-Hofstadter model strongly irradiated by light linearly polarized along the extra dimension in the superspace. In particular, the rescaling of the quasi-disorder $\Delta$ which we observe in the 1D model corresponds to a rescaling of tunneling along that dimension. This provides a complementary picture of the destruction of localization we observe, which in the higher-dimensional space appears as coherent destruction of tunneling along the extra dimension~\cite{chang_berry_1996_DrivenAA}, causing the 2D square lattice to decompose into a set of decoupled  one-dimensional chains which cannot support localized modes due to the absence of disorder.  Besides providing an alternative perspective, the superspace picture can also be used~\footnote{See the explicit demonstration in the supplementary information.} to design modulation protocols which perfectly destroy localization in a generic bounded quasiperiodic system, by connecting to the concept of exact dynamic localization~\cite{dignam_conditions_2002_EDL, domachuk_dynamic_2002_EDL}.  
This higher-dimensional mapping extends the applicability of our results to other quasiperiodic systems~\cite{an_interactions_2021_GAAH, goblot_emergence_2020_IAAF} and also implies an interpretation of our results as the first observation of dynamic localization in a strongly driven Harper-Hofstadter model. 

%This higher-dimensional mapping implies that b
Because dynamic localization is coherent~\cite{zenesini_coherent_2009_DL_Coherent}, phasonic modulation can be used as a tool to reversibly and coherently control transport. To experimentally test this possibility we performed transport measurements for several different driving sequences during which phason modulation is turned on and off at different times during the course of an experiment. Figure 3 shows the results of these experiments, compared to evolution in a static primary lattice and in a static bichromatic lattice. In a bichromatic lattice subjected to continuous phason modulation, the system evolves in a delocalized way, with width growing nearly as fast as when the secondary lattice was entirely absent. 
This further supports the notion that the quasiperiodic potential effectively vanishes at these resonant amplitudes. For a modulation protocol where phasonic driving is present only for the first 0.5 s of the evolution, the width grows rapidly in accordance with the delocalized expectation until the drive ceases, at which point the system localizes and the width becomes static. This observation indicates that the drive-induced delocalization is not due to significant heating or interband excitation, but rather represents coherent control of the localization properties. Anderson-type localization requires wavepacket coherence, and dephasing across lattice sites generally leads to delocalization as the coherence is destroyed~\cite{rayanov_decohering_2013, skokos_nonequilibrium_2013}; the fact that the atoms re-localize when the phasonic modulation is turned off indicates that coherence is maintained throughout the experiment. Finally, if phasonic modulation is applied only during the middle 0.5 s of the sequence, the width evolves in a localized way before the drive, then grows rapidly during the delocalized segment, then ceases to grow when the drive is removed. The last two coherent control protocols result in an identical width at the end of the experiment despite their different modulation histories. Together these results clearly demonstrate that phasonic driving can reversibly and coherently control a localization quantum phase transition. 

\begin{figure}[h]
	\centering
	\includegraphics[width=\columnwidth]{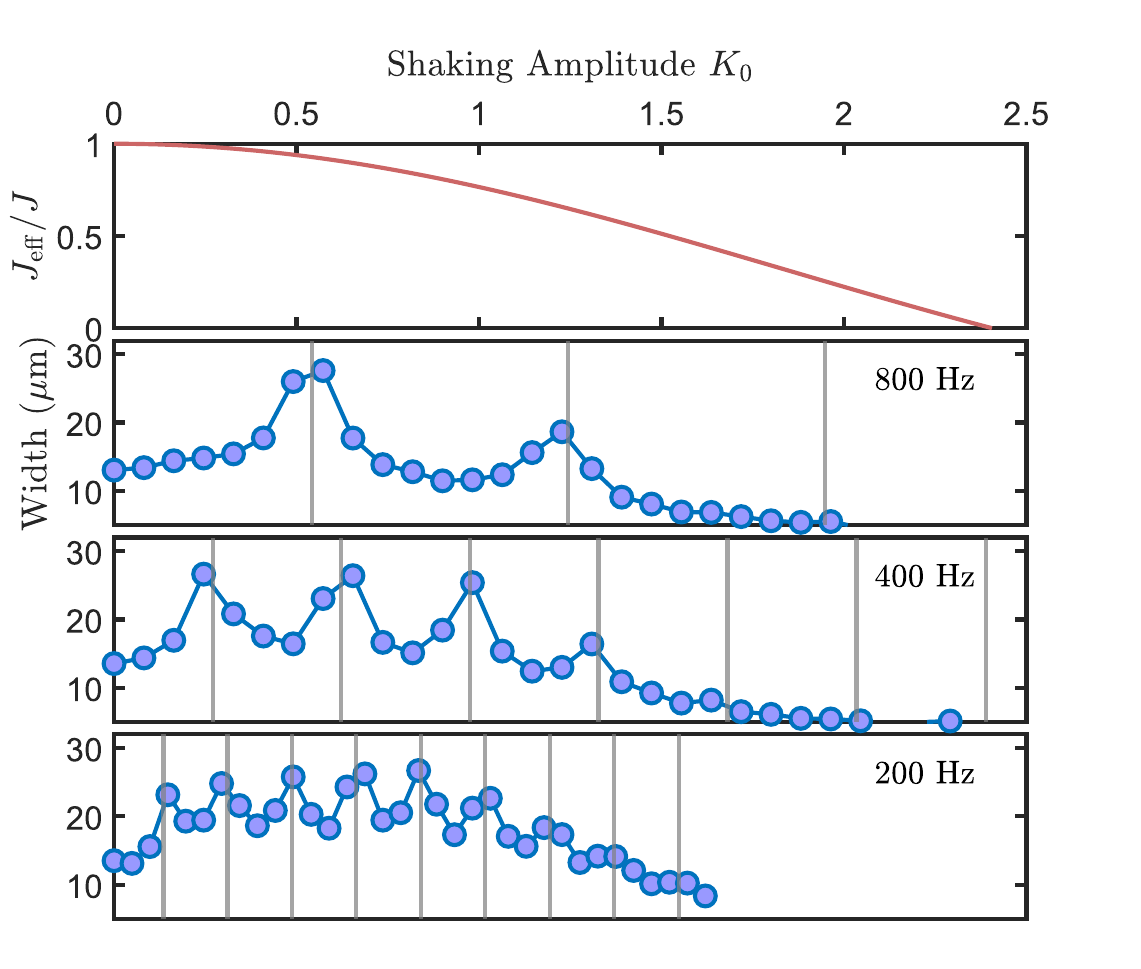} 
	\caption{Interplay between dynamic localization and Aubry-Andr\'e localization revealed by phase modulation of only the primary lattice. Top panel shows calculated effective tunneling strength as a function of modulation amplitude $K_0$. Lower panels show measured width of the density distribution after 1~s expansion in a bichromatic lattice with only the primary lattice shaken, for various modulation frequencies as indicated. Gray lines indicate theoretically expected values of zero effective quasi-disorder, as described in the text.
	}\label{Fig4}
\end{figure}

A new direction opened up by this capability is investigation of the interplay between dynamic localization (induced by a time-varying electric field) and Aubry-Andr\'e localization (induced by quasi-disorder)~\cite{PhysRevA.91.063825,Guzman-Silva.OptLett45.415}. Using the capabilities demonstrated above, both these types of localization can now be Floquet-tuned. In a final set of experiments we investigated this interplay, using phase modulation of just the primary lattice. For this experiment, we modified the setup of Fig.~1 so that the primary lattice could be phase modulated using paired acousto-optic modulators controlling the two beams comprising the lattice. In the co-moving frame of the shaken primary lattice, the atoms then experience both a phasonic modulation of the secondary lattice, which tunes Aubry-Andr\'e localization, and an alternating inertial force, which tunes the tunneling matrix element and drives dynamic localization. In the higher-dimensional picture described above, this corresponds to irradiation with elliptically polarized light.

Figure 4 shows the results of these experiments. The top panel shows the calculated effective tunneling matrix element as a function of the shaking amplitude $K_0=\pi \Delta\nu_{\rm{max}}/4 f_{\rm{r}}$, where $\Delta\nu_{\rm{max}}$ is the frequency modulation amplitude for the modulated primary lattice beam and $f_{\rm{r}}=E_{\rm{r}}/2 \pi\hbar$ as defined in \cite{eckardt_superfluid-insulator_2005,arimondo-shakenlattice}. The alternating inertial force rescales tunneling, controlling the overall expansion dynamics and leading to dynamic localization at $K_0\simeq 2.4$. However, since the modulation of the primary lattice also gives rise to phasonic driving, Aubry-Andr\'e delocalization can compete with this overall localizing trend, as modulation of the secondary lattice in the co-moving frame rescales the secondary lattice depth by the factor $J_0(2k_SA)$.  Since the phasonic modulation amplitude $A$ is related to $K_0$ by $A=(\lambda_p f_r/ \pi^2 f) K_0$, the same $K_0$ can correspond to different shaking amplitudes of the secondary lattice depending on the frequency of the phase modulation. The bottom panels of Fig. 4 show the late-time density distribution width after expansion in a bichromatic lattice with the primary lattice shaken at different frequencies $f$.  While all of the panels in Fig. 4 share the same $K_0$ axis, $A$ depends on the drive frequency. At each drive frequency, we observe an array of delocalizing peaks superimposed upon the overall trend towards dynamic localization as $K_0$ increases. The arrangement of delocalized peaks varies with the drive frequency, and the position of all observed delocalized peaks matches well to zeros of $J_0(2k_SA)$, indicated by gray lines in each panel. This close match to fit-parameter-free theory provides strong support for the interpretation of these delocalized peaks as being due to phasonic rescaling of the secondary lattice depth to zero in the comoving frame.

These results highlight the rich interplay between two distinct forms of localization, and open up additional possibilities. While in this work both phenomena were tuned with a single lattice modulation, a full exploration of the phase diagram of matter subjected to both dynamic localization and Aubry-Andr\'e localization, as envisioned in \cite{HolthausChapter}, would require separately controlling dynamic localization and the rescaling of the secondary lattice.

%CONCLUSION AND FUTURE DIRECTION
In summary, in this work we have demonstrated experimentally, and confirmed theoretically, that phasonic modulation in a quasicrystal can coherently control transport and reversibly tune across a localization-delocalization quantum phase transition.
We have shown that these results can be interpreted as manifestations of dynamic localization in the higher-dimensional lattice associated with the quasicrystal, opening up a pathway to simulation of strongly driven quantum Hall systems~\cite{arlinghaus_PhysRevA.81.063612_strong_field, wackerl_driven_2019_DrivenQHE, zhao_floquet_2022_DrivenQHE}. Combining both phasonic and dipolar driving would allow for complete control of the polarization (linear, elliptical, or circular) of the driving radiation which appears in the superspace, enabling quantum simulation of laser-irradiated integer quantum Hall systems with tunable incident polarization. The interplay between topology and modulation would be a natural direction for further investigation, as the minigap collapse which drives localization also signifies a topological transition~\cite{kraus_topological_2012_pumping}. 
Finally, the sign change of multiple Hamiltonian parameters across Bessel zeros opens up the possibility to design a modulation protocol which reverses the direction of time. 

%\clearpage
\begin{acknowledgments}
%FUNDING ACKNOWLEDGEMENTS STILL NEED THOUGHT
We acknowledge support from the Air Force Office of Scientific Research (FA9550-20-1-0240), the NSF QLCI program (OMA-2016245), the NSF Quantum Foundry (DMR-1906325), and the Army Research Office (MURI W911NF1710323). AE acknowledges funding by DFG via the Research Unit FOR 2414, under project No. 277974659. The optical lattices used herein were developed in work supported by the U.S. Department of Energy, Office of Science, National Quantum Information Science Research Centers, Quantum Science Center.
%DOE WANTS: QSC-led but with other funders: “This work was primarily led and supported by the U.S. Department of Energy, Office of Science, National Quantum Information Science Research Centers, Quantum Science Center (parts of work supported, people supported). [follow this with text for other funders]”
\end{acknowledgments}

\bibliographystyle{apsrev4-2}
\bibliography{refs}

\begin{thebibliography}{37}%
\makeatletter
\providecommand \@ifxundefined [1]{%
 \@ifx{#1\undefined}
}%
\providecommand \@ifnum [1]{%
 \ifnum #1\expandafter \@firstoftwo
 \else \expandafter \@secondoftwo
 \fi
}%
\providecommand \@ifx [1]{%
 \ifx #1\expandafter \@firstoftwo
 \else \expandafter \@secondoftwo
 \fi
}%
\providecommand \natexlab [1]{#1}%
\providecommand \enquote  [1]{``#1''}%
\providecommand \bibnamefont  [1]{#1}%
\providecommand \bibfnamefont [1]{#1}%
\providecommand \citenamefont [1]{#1}%
\providecommand \href@noop [0]{\@secondoftwo}%
\providecommand \href [0]{\begingroup \@sanitize@url \@href}%
\providecommand \@href[1]{\@@startlink{#1}\@@href}%
\providecommand \@@href[1]{\endgroup#1\@@endlink}%
\providecommand \@sanitize@url [0]{\catcode `\\12\catcode `\$12\catcode
  `\&12\catcode `\#12\catcode `\^12\catcode `\_12\catcode `\%12\relax}%
\providecommand \@@startlink[1]{}%
\providecommand \@@endlink[0]{}%
\providecommand \url  [0]{\begingroup\@sanitize@url \@url }%
\providecommand \@url [1]{\endgroup\@href {#1}{\urlprefix }}%
\providecommand \urlprefix  [0]{URL }%
\providecommand \Eprint [0]{\href }%
\providecommand \doibase [0]{https://doi.org/}%
\providecommand \selectlanguage [0]{\@gobble}%
\providecommand \bibinfo  [0]{\@secondoftwo}%
\providecommand \bibfield  [0]{\@secondoftwo}%
\providecommand \translation [1]{[#1]}%
\providecommand \BibitemOpen [0]{}%
\providecommand \bibitemStop [0]{}%
\providecommand \bibitemNoStop [0]{.\EOS\space}%
\providecommand \EOS [0]{\spacefactor3000\relax}%
\providecommand \BibitemShut  [1]{\csname bibitem#1\endcsname}%
\let\auto@bib@innerbib\@empty
%</preamble>
\bibitem [{\citenamefont {Eckardt}(2017)}]{RevModPhys.89.011004}%
  \BibitemOpen
  \bibfield  {author} {\bibinfo {author} {\bibfnamefont {A.}~\bibnamefont
  {Eckardt}},\ }\href {https://doi.org/10.1103/RevModPhys.89.011004} {\bibfield
   {journal} {\bibinfo  {journal} {Rev. Mod. Phys.}\ }\textbf {\bibinfo
  {volume} {89}},\ \bibinfo {pages} {011004} (\bibinfo {year}
  {2017})}\BibitemShut {NoStop}%
\bibitem [{\citenamefont {Lignier}\ \emph {et~al.}(2007)\citenamefont
  {Lignier}, \citenamefont {Sias}, \citenamefont {Ciampini}, \citenamefont
  {Singh}, \citenamefont {Zenesini}, \citenamefont {Morsch},\ and\
  \citenamefont {Arimondo}}]{arimondo-shakenlattice}%
  \BibitemOpen
  \bibfield  {author} {\bibinfo {author} {\bibfnamefont {H.}~\bibnamefont
  {Lignier}}, \bibinfo {author} {\bibfnamefont {C.}~\bibnamefont {Sias}},
  \bibinfo {author} {\bibfnamefont {D.}~\bibnamefont {Ciampini}}, \bibinfo
  {author} {\bibfnamefont {Y.}~\bibnamefont {Singh}}, \bibinfo {author}
  {\bibfnamefont {A.}~\bibnamefont {Zenesini}}, \bibinfo {author}
  {\bibfnamefont {O.}~\bibnamefont {Morsch}},\ and\ \bibinfo {author}
  {\bibfnamefont {E.}~\bibnamefont {Arimondo}},\ }\href
  {https://doi.org/10.1103/PhysRevLett.99.220403} {\bibfield  {journal}
  {\bibinfo  {journal} {Phys. Rev. Lett.}\ }\textbf {\bibinfo {volume} {99}},\
  \bibinfo {pages} {220403} (\bibinfo {year} {2007})}\BibitemShut {NoStop}%
\bibitem [{\citenamefont {Kierig}\ \emph {et~al.}(2008)\citenamefont {Kierig},
  \citenamefont {Schnorrberger}, \citenamefont {Schietinger}, \citenamefont
  {Tomkovic},\ and\ \citenamefont {Oberthaler}}]{oberthaler-drivendoublewells}%
  \BibitemOpen
  \bibfield  {author} {\bibinfo {author} {\bibfnamefont {E.}~\bibnamefont
  {Kierig}}, \bibinfo {author} {\bibfnamefont {U.}~\bibnamefont
  {Schnorrberger}}, \bibinfo {author} {\bibfnamefont {A.}~\bibnamefont
  {Schietinger}}, \bibinfo {author} {\bibfnamefont {J.}~\bibnamefont
  {Tomkovic}},\ and\ \bibinfo {author} {\bibfnamefont {M.~K.}\ \bibnamefont
  {Oberthaler}},\ }\href {https://doi.org/10.1103/PhysRevLett.100.190405}
  {\bibfield  {journal} {\bibinfo  {journal} {Phys. Rev. Lett.}\ }\textbf
  {\bibinfo {volume} {100}},\ \bibinfo {pages} {190405} (\bibinfo {year}
  {2008})}\BibitemShut {NoStop}%
\bibitem [{\citenamefont {Dunlap}\ and\ \citenamefont
  {Kenkre}(1986)}]{dunlap-dynloctheory}%
  \BibitemOpen
  \bibfield  {author} {\bibinfo {author} {\bibfnamefont {D.~H.}\ \bibnamefont
  {Dunlap}}\ and\ \bibinfo {author} {\bibfnamefont {V.~M.}\ \bibnamefont
  {Kenkre}},\ }\href {https://doi.org/10.1103/PhysRevB.34.3625} {\bibfield
  {journal} {\bibinfo  {journal} {Phys. Rev. B}\ }\textbf {\bibinfo {volume}
  {34}},\ \bibinfo {pages} {3625} (\bibinfo {year} {1986})}\BibitemShut
  {NoStop}%
\bibitem [{\citenamefont {Holthaus}(1992)}]{holthaus-dynloctheory_DL}%
  \BibitemOpen
  \bibfield  {author} {\bibinfo {author} {\bibfnamefont {M.}~\bibnamefont
  {Holthaus}},\ }\href {https://doi.org/10.1103/PhysRevLett.69.351} {\bibfield
  {journal} {\bibinfo  {journal} {Phys. Rev. Lett.}\ }\textbf {\bibinfo
  {volume} {69}},\ \bibinfo {pages} {351} (\bibinfo {year} {1992})}\BibitemShut
  {NoStop}%
\bibitem [{\citenamefont {Zenesini}\ \emph
  {et~al.}(2009{\natexlab{a}})\citenamefont {Zenesini}, \citenamefont
  {Lignier}, \citenamefont {Ciampini}, \citenamefont {Morsch},\ and\
  \citenamefont {Arimondo}}]{arimondo-coherentcontrol}%
  \BibitemOpen
  \bibfield  {author} {\bibinfo {author} {\bibfnamefont {A.}~\bibnamefont
  {Zenesini}}, \bibinfo {author} {\bibfnamefont {H.}~\bibnamefont {Lignier}},
  \bibinfo {author} {\bibfnamefont {D.}~\bibnamefont {Ciampini}}, \bibinfo
  {author} {\bibfnamefont {O.}~\bibnamefont {Morsch}},\ and\ \bibinfo {author}
  {\bibfnamefont {E.}~\bibnamefont {Arimondo}},\ }\href
  {https://doi.org/10.1103/PhysRevLett.102.100403} {\bibfield  {journal}
  {\bibinfo  {journal} {Phys. Rev. Lett.}\ }\textbf {\bibinfo {volume} {102}},\
  \bibinfo {pages} {100403} (\bibinfo {year} {2009}{\natexlab{a}})}\BibitemShut
  {NoStop}%
\bibitem [{\citenamefont {Rajagopal}\ \emph {et~al.}(2019)\citenamefont
  {Rajagopal}, \citenamefont {Shimasaki}, \citenamefont {Dotti}, \citenamefont
  {Ra\ifmmode \check{c}\else \v{c}\fi{}i\ifmmode~\bar{u}\else \={u}\fi{}nas},
  \citenamefont {Senaratne}, \citenamefont {Anisimovas}, \citenamefont
  {Eckardt},\ and\ \citenamefont {Weld}}]{shankari-phasonicspectroscopy}%
  \BibitemOpen
  \bibfield  {author} {\bibinfo {author} {\bibfnamefont {S.~V.}\ \bibnamefont
  {Rajagopal}}, \bibinfo {author} {\bibfnamefont {T.}~\bibnamefont
  {Shimasaki}}, \bibinfo {author} {\bibfnamefont {P.}~\bibnamefont {Dotti}},
  \bibinfo {author} {\bibfnamefont {M.}~\bibnamefont {Ra\ifmmode \check{c}\else
  \v{c}\fi{}i\ifmmode~\bar{u}\else \={u}\fi{}nas}}, \bibinfo {author}
  {\bibfnamefont {R.}~\bibnamefont {Senaratne}}, \bibinfo {author}
  {\bibfnamefont {E.}~\bibnamefont {Anisimovas}}, \bibinfo {author}
  {\bibfnamefont {A.}~\bibnamefont {Eckardt}},\ and\ \bibinfo {author}
  {\bibfnamefont {D.~M.}\ \bibnamefont {Weld}},\ }\href
  {https://doi.org/10.1103/PhysRevLett.123.223201} {\bibfield  {journal}
  {\bibinfo  {journal} {Phys. Rev. Lett.}\ }\textbf {\bibinfo {volume} {123}},\
  \bibinfo {pages} {223201} (\bibinfo {year} {2019})}\BibitemShut {NoStop}%
\bibitem [{\citenamefont {Kraus}\ and\ \citenamefont
  {Zilberberg}(2016)}]{kraus_quasiperiodicity_2016}%
  \BibitemOpen
  \bibfield  {author} {\bibinfo {author} {\bibfnamefont {Y.~E.}\ \bibnamefont
  {Kraus}}\ and\ \bibinfo {author} {\bibfnamefont {O.}~\bibnamefont
  {Zilberberg}},\ }\href {https://doi.org/10.1038/nphys3784} {\bibfield
  {journal} {\bibinfo  {journal} {Nature Physics}\ }\textbf {\bibinfo {volume}
  {12}},\ \bibinfo {pages} {624} (\bibinfo {year} {2016})}\BibitemShut
  {NoStop}%
\bibitem [{\citenamefont {Zilberberg}(2021)}]{zilberberg_topology_2021}%
  \BibitemOpen
  \bibfield  {author} {\bibinfo {author} {\bibfnamefont {O.}~\bibnamefont
  {Zilberberg}},\ }\href {https://doi.org/10.1364/OME.416552} {\bibfield
  {journal} {\bibinfo  {journal} {Optical Materials Express}\ }\textbf
  {\bibinfo {volume} {11}},\ \bibinfo {pages} {1143} (\bibinfo {year}
  {2021})}\BibitemShut {NoStop}%
\bibitem [{\citenamefont {Fan}\ and\ \citenamefont
  {Huang}(2022)}]{fan_topological_2022}%
  \BibitemOpen
  \bibfield  {author} {\bibinfo {author} {\bibfnamefont {J.}~\bibnamefont
  {Fan}}\ and\ \bibinfo {author} {\bibfnamefont {H.}~\bibnamefont {Huang}},\
  }\href {https://doi.org/10.1007/s11467-021-1100-y} {\bibfield  {journal}
  {\bibinfo  {journal} {Frontiers of Physics}\ }\textbf {\bibinfo {volume}
  {17}},\ \bibinfo {pages} {13203} (\bibinfo {year} {2022})}\BibitemShut
  {NoStop}%
\bibitem [{\citenamefont {Harper}(1955)}]{harper_single_1955}%
  \BibitemOpen
  \bibfield  {author} {\bibinfo {author} {\bibfnamefont {P.~G.}\ \bibnamefont
  {Harper}},\ }\href {https://doi.org/10.1088/0370-1298/68/10/304} {\bibfield
  {journal} {\bibinfo  {journal} {Proceedings of the Physical Society. Section
  A}\ }\textbf {\bibinfo {volume} {68}},\ \bibinfo {pages} {874} (\bibinfo
  {year} {1955})}\BibitemShut {NoStop}%
\bibitem [{\citenamefont {Kraus}\ \emph {et~al.}(2012)\citenamefont {Kraus},
  \citenamefont {Lahini}, \citenamefont {Ringel}, \citenamefont {Verbin},\ and\
  \citenamefont {Zilberberg}}]{kraus_topological_2012_pumping}%
  \BibitemOpen
  \bibfield  {author} {\bibinfo {author} {\bibfnamefont {Y.~E.}\ \bibnamefont
  {Kraus}}, \bibinfo {author} {\bibfnamefont {Y.}~\bibnamefont {Lahini}},
  \bibinfo {author} {\bibfnamefont {Z.}~\bibnamefont {Ringel}}, \bibinfo
  {author} {\bibfnamefont {M.}~\bibnamefont {Verbin}},\ and\ \bibinfo {author}
  {\bibfnamefont {O.}~\bibnamefont {Zilberberg}},\ }\href
  {https://doi.org/10.1103/PhysRevLett.109.106402} {\bibfield  {journal}
  {\bibinfo  {journal} {Physical Review Letters}\ }\textbf {\bibinfo {volume}
  {109}},\ \bibinfo {pages} {106402} (\bibinfo {year} {2012})}\BibitemShut
  {NoStop}%
\bibitem [{\citenamefont {Aubry}\ and\ \citenamefont
  {Andr{\'e}}(1980)}]{aubry1980analyticity}%
  \BibitemOpen
  \bibfield  {author} {\bibinfo {author} {\bibfnamefont {S.}~\bibnamefont
  {Aubry}}\ and\ \bibinfo {author} {\bibfnamefont {G.}~\bibnamefont
  {Andr{\'e}}},\ }\href@noop {} {\bibfield  {journal} {\bibinfo  {journal}
  {Ann. Israel Phys. Soc}\ }\textbf {\bibinfo {volume} {3}},\ \bibinfo {pages}
  {18} (\bibinfo {year} {1980})}\BibitemShut {NoStop}%
\bibitem [{\citenamefont {Roati}\ \emph {et~al.}(2008)\citenamefont {Roati},
  \citenamefont {D’Errico}, \citenamefont {Fallani}, \citenamefont {Fattori},
  \citenamefont {Fort}, \citenamefont {Zaccanti}, \citenamefont {Modugno},
  \citenamefont {Modugno},\ and\ \citenamefont
  {Inguscio}}]{roati_anderson_2008_AAExp}%
  \BibitemOpen
  \bibfield  {author} {\bibinfo {author} {\bibfnamefont {G.}~\bibnamefont
  {Roati}}, \bibinfo {author} {\bibfnamefont {C.}~\bibnamefont {D’Errico}},
  \bibinfo {author} {\bibfnamefont {L.}~\bibnamefont {Fallani}}, \bibinfo
  {author} {\bibfnamefont {M.}~\bibnamefont {Fattori}}, \bibinfo {author}
  {\bibfnamefont {C.}~\bibnamefont {Fort}}, \bibinfo {author} {\bibfnamefont
  {M.}~\bibnamefont {Zaccanti}}, \bibinfo {author} {\bibfnamefont
  {G.}~\bibnamefont {Modugno}}, \bibinfo {author} {\bibfnamefont
  {M.}~\bibnamefont {Modugno}},\ and\ \bibinfo {author} {\bibfnamefont
  {M.}~\bibnamefont {Inguscio}},\ }\href {https://doi.org/10.1038/nature07071}
  {\bibfield  {journal} {\bibinfo  {journal} {Nature}\ }\textbf {\bibinfo
  {volume} {453}},\ \bibinfo {pages} {895} (\bibinfo {year}
  {2008})}\BibitemShut {NoStop}%
\bibitem [{\citenamefont {Singh}\ \emph {et~al.}(2015)\citenamefont {Singh},
  \citenamefont {Saha}, \citenamefont {Parameswaran},\ and\ \citenamefont
  {Weld}}]{kevinQCpaper}%
  \BibitemOpen
  \bibfield  {author} {\bibinfo {author} {\bibfnamefont {K.}~\bibnamefont
  {Singh}}, \bibinfo {author} {\bibfnamefont {K.}~\bibnamefont {Saha}},
  \bibinfo {author} {\bibfnamefont {S.~A.}\ \bibnamefont {Parameswaran}},\ and\
  \bibinfo {author} {\bibfnamefont {D.~M.}\ \bibnamefont {Weld}},\ }\href
  {https://doi.org/10.1103/PhysRevA.92.063426} {\bibfield  {journal} {\bibinfo
  {journal} {Phys. Rev. A}\ }\textbf {\bibinfo {volume} {92}},\ \bibinfo
  {pages} {063426} (\bibinfo {year} {2015})}\BibitemShut {NoStop}%
\bibitem [{\citenamefont {Viebahn}\ \emph {et~al.}(2019)\citenamefont
  {Viebahn}, \citenamefont {Sbroscia}, \citenamefont {Carter}, \citenamefont
  {Yu},\ and\ \citenamefont {Schneider}}]{SchneiderQuasicrystalPRL1}%
  \BibitemOpen
  \bibfield  {author} {\bibinfo {author} {\bibfnamefont {K.}~\bibnamefont
  {Viebahn}}, \bibinfo {author} {\bibfnamefont {M.}~\bibnamefont {Sbroscia}},
  \bibinfo {author} {\bibfnamefont {E.}~\bibnamefont {Carter}}, \bibinfo
  {author} {\bibfnamefont {J.-C.}\ \bibnamefont {Yu}},\ and\ \bibinfo {author}
  {\bibfnamefont {U.}~\bibnamefont {Schneider}},\ }\href
  {https://doi.org/10.1103/PhysRevLett.122.110404} {\bibfield  {journal}
  {\bibinfo  {journal} {Phys. Rev. Lett.}\ }\textbf {\bibinfo {volume} {122}},\
  \bibinfo {pages} {110404} (\bibinfo {year} {2019})}\BibitemShut {NoStop}%
\bibitem [{\citenamefont {Sbroscia}\ \emph {et~al.}(2020)\citenamefont
  {Sbroscia}, \citenamefont {Viebahn}, \citenamefont {Carter}, \citenamefont
  {Yu}, \citenamefont {Gaunt},\ and\ \citenamefont
  {Schneider}}]{SchneiderQuasicrystalPRL2}%
  \BibitemOpen
  \bibfield  {author} {\bibinfo {author} {\bibfnamefont {M.}~\bibnamefont
  {Sbroscia}}, \bibinfo {author} {\bibfnamefont {K.}~\bibnamefont {Viebahn}},
  \bibinfo {author} {\bibfnamefont {E.}~\bibnamefont {Carter}}, \bibinfo
  {author} {\bibfnamefont {J.-C.}\ \bibnamefont {Yu}}, \bibinfo {author}
  {\bibfnamefont {A.}~\bibnamefont {Gaunt}},\ and\ \bibinfo {author}
  {\bibfnamefont {U.}~\bibnamefont {Schneider}},\ }\href
  {https://doi.org/10.1103/PhysRevLett.125.200604} {\bibfield  {journal}
  {\bibinfo  {journal} {Phys. Rev. Lett.}\ }\textbf {\bibinfo {volume} {125}},\
  \bibinfo {pages} {200604} (\bibinfo {year} {2020})}\BibitemShut {NoStop}%
\bibitem [{\citenamefont {Jitomirskaya}(1999)}]{jyotirmskaya}%
  \BibitemOpen
  \bibfield  {author} {\bibinfo {author} {\bibfnamefont {S.~Y.}\ \bibnamefont
  {Jitomirskaya}},\ }\href {http://www.jstor.org/stable/121066} {\bibfield
  {journal} {\bibinfo  {journal} {Annals of Mathematics}\ }\textbf {\bibinfo
  {volume} {150}},\ \bibinfo {pages} {1159} (\bibinfo {year}
  {1999})}\BibitemShut {NoStop}%
\bibitem [{\citenamefont {Sun}\ and\ \citenamefont
  {Eckardt}(2020)}]{eckardt-optimalfreq}%
  \BibitemOpen
  \bibfield  {author} {\bibinfo {author} {\bibfnamefont {G.}~\bibnamefont
  {Sun}}\ and\ \bibinfo {author} {\bibfnamefont {A.}~\bibnamefont {Eckardt}},\
  }\href {https://doi.org/10.1103/PhysRevResearch.2.013241} {\bibfield
  {journal} {\bibinfo  {journal} {Phys. Rev. Research}\ }\textbf {\bibinfo
  {volume} {2}},\ \bibinfo {pages} {013241} (\bibinfo {year}
  {2020})}\BibitemShut {NoStop}%
\bibitem [{Note1()}]{Note1}%
  \BibitemOpen
  \bibinfo {note} {See supplementary information.}\BibitemShut {Stop}%
\bibitem [{\citenamefont {Hofstadter}(1976)}]{hofstadter_energy_1976_HH}%
  \BibitemOpen
  \bibfield  {author} {\bibinfo {author} {\bibfnamefont {D.~R.}\ \bibnamefont
  {Hofstadter}},\ }\href {https://doi.org/10.1103/PhysRevB.14.2239} {\bibfield
  {journal} {\bibinfo  {journal} {Physical Review B}\ }\textbf {\bibinfo
  {volume} {14}},\ \bibinfo {pages} {2239} (\bibinfo {year}
  {1976})}\BibitemShut {NoStop}%
\bibitem [{\citenamefont {Chang}\ and\ \citenamefont
  {Niu}(1996)}]{chang_berry_1996_DrivenAA}%
  \BibitemOpen
  \bibfield  {author} {\bibinfo {author} {\bibfnamefont {M.-C.}\ \bibnamefont
  {Chang}}\ and\ \bibinfo {author} {\bibfnamefont {Q.}~\bibnamefont {Niu}},\
  }\href {https://doi.org/10.1103/PhysRevB.53.7010} {\bibfield  {journal}
  {\bibinfo  {journal} {Physical Review B}\ }\textbf {\bibinfo {volume} {53}},\
  \bibinfo {pages} {7010} (\bibinfo {year} {1996})}\BibitemShut {NoStop}%
\bibitem [{Note2()}]{Note2}%
  \BibitemOpen
  \bibinfo {note} {See the explicit demonstration in the supplementary
  information.}\BibitemShut {Stop}%
\bibitem [{\citenamefont {Dignam}\ and\ \citenamefont
  {De~Sterke}(2002)}]{dignam_conditions_2002_EDL}%
  \BibitemOpen
  \bibfield  {author} {\bibinfo {author} {\bibfnamefont {M.~M.}\ \bibnamefont
  {Dignam}}\ and\ \bibinfo {author} {\bibfnamefont {C.~M.}\ \bibnamefont
  {De~Sterke}},\ }\href {https://doi.org/10.1103/PhysRevLett.88.046806}
  {\bibfield  {journal} {\bibinfo  {journal} {Physical Review Letters}\
  }\textbf {\bibinfo {volume} {88}},\ \bibinfo {pages} {046806} (\bibinfo
  {year} {2002})}\BibitemShut {NoStop}%
\bibitem [{\citenamefont {Domachuk}\ \emph {et~al.}(2002)\citenamefont
  {Domachuk}, \citenamefont {Martijn De~Sterke}, \citenamefont {Wan},\ and\
  \citenamefont {Dignam}}]{domachuk_dynamic_2002_EDL}%
  \BibitemOpen
  \bibfield  {author} {\bibinfo {author} {\bibfnamefont {P.}~\bibnamefont
  {Domachuk}}, \bibinfo {author} {\bibfnamefont {C.}~\bibnamefont {Martijn
  De~Sterke}}, \bibinfo {author} {\bibfnamefont {J.}~\bibnamefont {Wan}},\ and\
  \bibinfo {author} {\bibfnamefont {M.~M.}\ \bibnamefont {Dignam}},\ }\href
  {https://doi.org/10.1103/PhysRevB.66.165313} {\bibfield  {journal} {\bibinfo
  {journal} {Physical Review B}\ }\textbf {\bibinfo {volume} {66}},\ \bibinfo
  {pages} {165313} (\bibinfo {year} {2002})}\BibitemShut {NoStop}%
\bibitem [{\citenamefont {An}\ \emph {et~al.}(2021)\citenamefont {An},
  \citenamefont {Padavić}, \citenamefont {Meier}, \citenamefont {Hegde},
  \citenamefont {Ganeshan}, \citenamefont {Pixley}, \citenamefont
  {Vishveshwara},\ and\ \citenamefont {Gadway}}]{an_interactions_2021_GAAH}%
  \BibitemOpen
  \bibfield  {author} {\bibinfo {author} {\bibfnamefont {F.~A.}\ \bibnamefont
  {An}}, \bibinfo {author} {\bibfnamefont {K.}~\bibnamefont {Padavić}},
  \bibinfo {author} {\bibfnamefont {E.~J.}\ \bibnamefont {Meier}}, \bibinfo
  {author} {\bibfnamefont {S.}~\bibnamefont {Hegde}}, \bibinfo {author}
  {\bibfnamefont {S.}~\bibnamefont {Ganeshan}}, \bibinfo {author}
  {\bibfnamefont {J.}~\bibnamefont {Pixley}}, \bibinfo {author} {\bibfnamefont
  {S.}~\bibnamefont {Vishveshwara}},\ and\ \bibinfo {author} {\bibfnamefont
  {B.}~\bibnamefont {Gadway}},\ }\href
  {https://doi.org/10.1103/PhysRevLett.126.040603} {\bibfield  {journal}
  {\bibinfo  {journal} {Physical Review Letters}\ }\textbf {\bibinfo {volume}
  {126}},\ \bibinfo {pages} {040603} (\bibinfo {year} {2021})}\BibitemShut
  {NoStop}%
\bibitem [{\citenamefont {Goblot}\ \emph {et~al.}(2020)\citenamefont {Goblot},
  \citenamefont {Štrkalj}, \citenamefont {Pernet}, \citenamefont {Lado},
  \citenamefont {Dorow}, \citenamefont {Lemaître}, \citenamefont {Le~Gratiet},
  \citenamefont {Harouri}, \citenamefont {Sagnes}, \citenamefont {Ravets},
  \citenamefont {Amo}, \citenamefont {Bloch},\ and\ \citenamefont
  {Zilberberg}}]{goblot_emergence_2020_IAAF}%
  \BibitemOpen
  \bibfield  {author} {\bibinfo {author} {\bibfnamefont {V.}~\bibnamefont
  {Goblot}}, \bibinfo {author} {\bibfnamefont {A.}~\bibnamefont {Štrkalj}},
  \bibinfo {author} {\bibfnamefont {N.}~\bibnamefont {Pernet}}, \bibinfo
  {author} {\bibfnamefont {J.~L.}\ \bibnamefont {Lado}}, \bibinfo {author}
  {\bibfnamefont {C.}~\bibnamefont {Dorow}}, \bibinfo {author} {\bibfnamefont
  {A.}~\bibnamefont {Lemaître}}, \bibinfo {author} {\bibfnamefont
  {L.}~\bibnamefont {Le~Gratiet}}, \bibinfo {author} {\bibfnamefont
  {A.}~\bibnamefont {Harouri}}, \bibinfo {author} {\bibfnamefont
  {I.}~\bibnamefont {Sagnes}}, \bibinfo {author} {\bibfnamefont
  {S.}~\bibnamefont {Ravets}}, \bibinfo {author} {\bibfnamefont
  {A.}~\bibnamefont {Amo}}, \bibinfo {author} {\bibfnamefont {J.}~\bibnamefont
  {Bloch}},\ and\ \bibinfo {author} {\bibfnamefont {O.}~\bibnamefont
  {Zilberberg}},\ }\href {https://doi.org/10.1038/s41567-020-0908-7} {\bibfield
   {journal} {\bibinfo  {journal} {Nature Physics}\ }\textbf {\bibinfo {volume}
  {16}},\ \bibinfo {pages} {832} (\bibinfo {year} {2020})}\BibitemShut
  {NoStop}%
\bibitem [{\citenamefont {Zenesini}\ \emph
  {et~al.}(2009{\natexlab{b}})\citenamefont {Zenesini}, \citenamefont
  {Lignier}, \citenamefont {Ciampini}, \citenamefont {Morsch},\ and\
  \citenamefont {Arimondo}}]{zenesini_coherent_2009_DL_Coherent}%
  \BibitemOpen
  \bibfield  {author} {\bibinfo {author} {\bibfnamefont {A.}~\bibnamefont
  {Zenesini}}, \bibinfo {author} {\bibfnamefont {H.}~\bibnamefont {Lignier}},
  \bibinfo {author} {\bibfnamefont {D.}~\bibnamefont {Ciampini}}, \bibinfo
  {author} {\bibfnamefont {O.}~\bibnamefont {Morsch}},\ and\ \bibinfo {author}
  {\bibfnamefont {E.}~\bibnamefont {Arimondo}},\ }\href
  {https://doi.org/10.1103/PhysRevLett.102.100403} {\bibfield  {journal}
  {\bibinfo  {journal} {Physical Review Letters}\ }\textbf {\bibinfo {volume}
  {102}},\ \bibinfo {pages} {100403} (\bibinfo {year}
  {2009}{\natexlab{b}})}\BibitemShut {NoStop}%
\bibitem [{\citenamefont {Rayanov}\ \emph {et~al.}(2013)\citenamefont
  {Rayanov}, \citenamefont {Radons},\ and\ \citenamefont
  {Flach}}]{rayanov_decohering_2013}%
  \BibitemOpen
  \bibfield  {author} {\bibinfo {author} {\bibfnamefont {K.}~\bibnamefont
  {Rayanov}}, \bibinfo {author} {\bibfnamefont {G.}~\bibnamefont {Radons}},\
  and\ \bibinfo {author} {\bibfnamefont {S.}~\bibnamefont {Flach}},\ }\href
  {https://doi.org/10.1103/PhysRevE.88.012901} {\bibfield  {journal} {\bibinfo
  {journal} {Physical Review E}\ }\textbf {\bibinfo {volume} {88}},\ \bibinfo
  {pages} {012901} (\bibinfo {year} {2013})}\BibitemShut {NoStop}%
\bibitem [{\citenamefont {Skokos}\ \emph {et~al.}(2013)\citenamefont {Skokos},
  \citenamefont {Gkolias},\ and\ \citenamefont
  {Flach}}]{skokos_nonequilibrium_2013}%
  \BibitemOpen
  \bibfield  {author} {\bibinfo {author} {\bibfnamefont {C.}~\bibnamefont
  {Skokos}}, \bibinfo {author} {\bibfnamefont {I.}~\bibnamefont {Gkolias}},\
  and\ \bibinfo {author} {\bibfnamefont {S.}~\bibnamefont {Flach}},\ }\href
  {https://doi.org/10.1103/PhysRevLett.111.064101} {\bibfield  {journal}
  {\bibinfo  {journal} {Physical Review Letters}\ }\textbf {\bibinfo {volume}
  {111}},\ \bibinfo {pages} {064101} (\bibinfo {year} {2013})}\BibitemShut
  {NoStop}%
\bibitem [{\citenamefont {Borovkova}\ \emph {et~al.}(2015)\citenamefont
  {Borovkova}, \citenamefont {Lobanov}, \citenamefont {Kartashov},
  \citenamefont {Vysloukh},\ and\ \citenamefont {Torner}}]{PhysRevA.91.063825}%
  \BibitemOpen
  \bibfield  {author} {\bibinfo {author} {\bibfnamefont {O.~V.}\ \bibnamefont
  {Borovkova}}, \bibinfo {author} {\bibfnamefont {V.~E.}\ \bibnamefont
  {Lobanov}}, \bibinfo {author} {\bibfnamefont {Y.~V.}\ \bibnamefont
  {Kartashov}}, \bibinfo {author} {\bibfnamefont {V.~A.}\ \bibnamefont
  {Vysloukh}},\ and\ \bibinfo {author} {\bibfnamefont {L.}~\bibnamefont
  {Torner}},\ }\href {https://doi.org/10.1103/PhysRevA.91.063825} {\bibfield
  {journal} {\bibinfo  {journal} {Phys. Rev. A}\ }\textbf {\bibinfo {volume}
  {91}},\ \bibinfo {pages} {063825} (\bibinfo {year} {2015})}\BibitemShut
  {NoStop}%
\bibitem [{\citenamefont {Guzman-Silva}\ \emph {et~al.}(2020)\citenamefont
  {Guzman-Silva}, \citenamefont {Heinrich}, \citenamefont {Biesenthal},
  \citenamefont {Kartashov},\ and\ \citenamefont
  {Szameit}}]{Guzman-Silva.OptLett45.415}%
  \BibitemOpen
  \bibfield  {author} {\bibinfo {author} {\bibfnamefont {D.}~\bibnamefont
  {Guzman-Silva}}, \bibinfo {author} {\bibfnamefont {M.}~\bibnamefont
  {Heinrich}}, \bibinfo {author} {\bibfnamefont {T.}~\bibnamefont
  {Biesenthal}}, \bibinfo {author} {\bibfnamefont {Y.~V.}\ \bibnamefont
  {Kartashov}},\ and\ \bibinfo {author} {\bibfnamefont {A.}~\bibnamefont
  {Szameit}},\ }\href {https://doi.org/10.1364/OL.380399} {\bibfield  {journal}
  {\bibinfo  {journal} {Opt. Lett.}\ }\textbf {\bibinfo {volume} {45}},\
  \bibinfo {pages} {415} (\bibinfo {year} {2020})}\BibitemShut {NoStop}%
\bibitem [{\citenamefont {Eckardt}\ \emph {et~al.}(2005)\citenamefont
  {Eckardt}, \citenamefont {Weiss},\ and\ \citenamefont
  {Holthaus}}]{eckardt_superfluid-insulator_2005}%
  \BibitemOpen
  \bibfield  {author} {\bibinfo {author} {\bibfnamefont {A.}~\bibnamefont
  {Eckardt}}, \bibinfo {author} {\bibfnamefont {C.}~\bibnamefont {Weiss}},\
  and\ \bibinfo {author} {\bibfnamefont {M.}~\bibnamefont {Holthaus}},\ }\href
  {https://doi.org/10.1103/PhysRevLett.95.260404} {\bibfield  {journal}
  {\bibinfo  {journal} {Physical Review Letters}\ }\textbf {\bibinfo {volume}
  {95}},\ \bibinfo {pages} {260404} (\bibinfo {year} {2005})}\BibitemShut
  {NoStop}%
\bibitem [{\citenamefont {Arlinghaus}\ \emph {et~al.}(2011)\citenamefont
  {Arlinghaus}, \citenamefont {Langemeyer},\ and\ \citenamefont
  {Holthaus}}]{HolthausChapter}%
  \BibitemOpen
  \bibfield  {author} {\bibinfo {author} {\bibfnamefont {S.}~\bibnamefont
  {Arlinghaus}}, \bibinfo {author} {\bibfnamefont {M.}~\bibnamefont
  {Langemeyer}},\ and\ \bibinfo {author} {\bibfnamefont {M.}~\bibnamefont
  {Holthaus}},\ }in\ \href@noop {} {\emph {\bibinfo {booktitle} {Dynamical
  Tunneling: Theory and Experiment}}},\ \bibinfo {editor} {edited by\ \bibinfo
  {editor} {\bibfnamefont {S.}~\bibnamefont {Keshavamurthy}}\ and\ \bibinfo
  {editor} {\bibfnamefont {P.}~\bibnamefont {Schlagheck}}}\ (\bibinfo
  {publisher} {CRC Press},\ \bibinfo {year} {2011})\ \bibinfo {edition} {1st}\
  ed.,\ Chap.~\bibinfo {chapter} {12}\BibitemShut {NoStop}%
\bibitem [{\citenamefont {Arlinghaus}\ and\ \citenamefont
  {Holthaus}(2010)}]{arlinghaus_PhysRevA.81.063612_strong_field}%
  \BibitemOpen
  \bibfield  {author} {\bibinfo {author} {\bibfnamefont {S.}~\bibnamefont
  {Arlinghaus}}\ and\ \bibinfo {author} {\bibfnamefont {M.}~\bibnamefont
  {Holthaus}},\ }\href {https://doi.org/10.1103/PhysRevA.81.063612} {\bibfield
  {journal} {\bibinfo  {journal} {Phys. Rev. A}\ }\textbf {\bibinfo {volume}
  {81}},\ \bibinfo {pages} {063612} (\bibinfo {year} {2010})}\BibitemShut
  {NoStop}%
\bibitem [{\citenamefont {Wackerl}\ \emph {et~al.}(2019)\citenamefont
  {Wackerl}, \citenamefont {Wenk},\ and\ \citenamefont
  {Schliemann}}]{wackerl_driven_2019_DrivenQHE}%
  \BibitemOpen
  \bibfield  {author} {\bibinfo {author} {\bibfnamefont {M.}~\bibnamefont
  {Wackerl}}, \bibinfo {author} {\bibfnamefont {P.}~\bibnamefont {Wenk}},\ and\
  \bibinfo {author} {\bibfnamefont {J.}~\bibnamefont {Schliemann}},\ }\href
  {https://doi.org/10.1103/PhysRevB.100.165411} {\bibfield  {journal} {\bibinfo
   {journal} {Physical Review B}\ }\textbf {\bibinfo {volume} {100}},\ \bibinfo
  {pages} {165411} (\bibinfo {year} {2019})}\BibitemShut {NoStop}%
\bibitem [{\citenamefont {Zhao}\ \emph {et~al.}(2022)\citenamefont {Zhao},
  \citenamefont {Chen}, \citenamefont {Tian},\ and\ \citenamefont
  {Du}}]{zhao_floquet_2022_DrivenQHE}%
  \BibitemOpen
  \bibfield  {author} {\bibinfo {author} {\bibfnamefont {M.}~\bibnamefont
  {Zhao}}, \bibinfo {author} {\bibfnamefont {Q.}~\bibnamefont {Chen}}, \bibinfo
  {author} {\bibfnamefont {X.-D.}\ \bibnamefont {Tian}},\ and\ \bibinfo
  {author} {\bibfnamefont {L.}~\bibnamefont {Du}},\ }\href
  {https://doi.org/10.1088/1751-8121/ac7488} {\bibfield  {journal} {\bibinfo
  {journal} {Journal of Physics A: Mathematical and Theoretical}\ }\textbf
  {\bibinfo {volume} {55}},\ \bibinfo {pages} {275003} (\bibinfo {year}
  {2022})},\ \bibinfo {note} {arXiv:2007.01071 [cond-mat]}\BibitemShut
  {NoStop}%
\end{thebibliography}%


\begin{thebibliography}{10}%
\makeatletter
\providecommand \@ifxundefined [1]{%
 \@ifx{#1\undefined}
}%
\providecommand \@ifnum [1]{%
 \ifnum #1\expandafter \@firstoftwo
 \else \expandafter \@secondoftwo
 \fi
}%
\providecommand \@ifx [1]{%
 \ifx #1\expandafter \@firstoftwo
 \else \expandafter \@secondoftwo
 \fi
}%
\providecommand \natexlab [1]{#1}%
\providecommand \enquote  [1]{``#1''}%
\providecommand \bibnamefont  [1]{#1}%
\providecommand \bibfnamefont [1]{#1}%
\providecommand \citenamefont [1]{#1}%
\providecommand \href@noop [0]{\@secondoftwo}%
\providecommand \href [0]{\begingroup \@sanitize@url \@href}%
\providecommand \@href[1]{\@@startlink{#1}\@@href}%
\providecommand \@@href[1]{\endgroup#1\@@endlink}%
\providecommand \@sanitize@url [0]{\catcode `\\12\catcode `\$12\catcode
  `\&12\catcode `\#12\catcode `\^12\catcode `\_12\catcode `\%12\relax}%
\providecommand \@@startlink[1]{}%
\providecommand \@@endlink[0]{}%
\providecommand \url  [0]{\begingroup\@sanitize@url \@url }%
\providecommand \@url [1]{\endgroup\@href {#1}{\urlprefix }}%
\providecommand \urlprefix  [0]{URL }%
\providecommand \Eprint [0]{\href }%
\providecommand \doibase [0]{https://doi.org/}%
\providecommand \selectlanguage [0]{\@gobble}%
\providecommand \bibinfo  [0]{\@secondoftwo}%
\providecommand \bibfield  [0]{\@secondoftwo}%
\providecommand \translation [1]{[#1]}%
\providecommand \BibitemOpen [0]{}%
\providecommand \bibitemStop [0]{}%
\providecommand \bibitemNoStop [0]{.\EOS\space}%
\providecommand \EOS [0]{\spacefactor3000\relax}%
\providecommand \BibitemShut  [1]{\csname bibitem#1\endcsname}%
\let\auto@bib@innerbib\@empty
%</preamble>
\bibitem [{\citenamefont {Ganeshan}\ \emph {et~al.}(2015)\citenamefont
  {Ganeshan}, \citenamefont {Pixley},\ and\ \citenamefont
  {Das~Sarma}}]{ganeshan_nearest_2015_GAAH}%
  \BibitemOpen
  \bibfield  {author} {\bibinfo {author} {\bibfnamefont {S.}~\bibnamefont
  {Ganeshan}}, \bibinfo {author} {\bibfnamefont {J.}~\bibnamefont {Pixley}},\
  and\ \bibinfo {author} {\bibfnamefont {S.}~\bibnamefont {Das~Sarma}},\ }\href
  {https://doi.org/10.1103/PhysRevLett.114.146601} {\bibfield  {journal}
  {\bibinfo  {journal} {Physical Review Letters}\ }\textbf {\bibinfo {volume}
  {114}},\ \bibinfo {pages} {146601} (\bibinfo {year} {2015})}\BibitemShut
  {NoStop}%
\bibitem [{\citenamefont {Devakul}\ and\ \citenamefont
  {Huse}(2017)}]{devakul_anderson_2017_Superlattice}%
  \BibitemOpen
  \bibfield  {author} {\bibinfo {author} {\bibfnamefont {T.}~\bibnamefont
  {Devakul}}\ and\ \bibinfo {author} {\bibfnamefont {D.~A.}\ \bibnamefont
  {Huse}},\ }\href {https://doi.org/10.1103/PhysRevB.96.214201} {\bibfield
  {journal} {\bibinfo  {journal} {Physical Review B}\ }\textbf {\bibinfo
  {volume} {96}},\ \bibinfo {pages} {214201} (\bibinfo {year}
  {2017})}\BibitemShut {NoStop}%
\bibitem [{\citenamefont {Dignam}\ and\ \citenamefont
  {De~Sterke}(2002)}]{dignam_conditions_2002_EDL}%
  \BibitemOpen
  \bibfield  {author} {\bibinfo {author} {\bibfnamefont {M.~M.}\ \bibnamefont
  {Dignam}}\ and\ \bibinfo {author} {\bibfnamefont {C.~M.}\ \bibnamefont
  {De~Sterke}},\ }\href {https://doi.org/10.1103/PhysRevLett.88.046806}
  {\bibfield  {journal} {\bibinfo  {journal} {Physical Review Letters}\
  }\textbf {\bibinfo {volume} {88}},\ \bibinfo {pages} {046806} (\bibinfo
  {year} {2002})}\BibitemShut {NoStop}%
\bibitem [{\citenamefont {Domachuk}\ \emph {et~al.}(2002)\citenamefont
  {Domachuk}, \citenamefont {Martijn De~Sterke}, \citenamefont {Wan},\ and\
  \citenamefont {Dignam}}]{domachuk_dynamic_2002_EDL}%
  \BibitemOpen
  \bibfield  {author} {\bibinfo {author} {\bibfnamefont {P.}~\bibnamefont
  {Domachuk}}, \bibinfo {author} {\bibfnamefont {C.}~\bibnamefont {Martijn
  De~Sterke}}, \bibinfo {author} {\bibfnamefont {J.}~\bibnamefont {Wan}},\ and\
  \bibinfo {author} {\bibfnamefont {M.~M.}\ \bibnamefont {Dignam}},\ }\href
  {https://doi.org/10.1103/PhysRevB.66.165313} {\bibfield  {journal} {\bibinfo
  {journal} {Physical Review B}\ }\textbf {\bibinfo {volume} {66}},\ \bibinfo
  {pages} {165313} (\bibinfo {year} {2002})}\BibitemShut {NoStop}%
\bibitem [{\citenamefont {Eckardt}\ \emph {et~al.}(2009)\citenamefont
  {Eckardt}, \citenamefont {Holthaus}, \citenamefont {Lignier}, \citenamefont
  {Zenesini}, \citenamefont {Ciampini}, \citenamefont {Morsch},\ and\
  \citenamefont {Arimondo}}]{eckardt_exploring_2009_DL}%
  \BibitemOpen
  \bibfield  {author} {\bibinfo {author} {\bibfnamefont {A.}~\bibnamefont
  {Eckardt}}, \bibinfo {author} {\bibfnamefont {M.}~\bibnamefont {Holthaus}},
  \bibinfo {author} {\bibfnamefont {H.}~\bibnamefont {Lignier}}, \bibinfo
  {author} {\bibfnamefont {A.}~\bibnamefont {Zenesini}}, \bibinfo {author}
  {\bibfnamefont {D.}~\bibnamefont {Ciampini}}, \bibinfo {author}
  {\bibfnamefont {O.}~\bibnamefont {Morsch}},\ and\ \bibinfo {author}
  {\bibfnamefont {E.}~\bibnamefont {Arimondo}},\ }\href
  {https://doi.org/10.1103/PhysRevA.79.013611} {\bibfield  {journal} {\bibinfo
  {journal} {Physical Review A}\ }\textbf {\bibinfo {volume} {79}},\ \bibinfo
  {pages} {013611} (\bibinfo {year} {2009})}\BibitemShut {NoStop}%
\bibitem [{\citenamefont {Joushaghani}\ \emph {et~al.}(2012)\citenamefont
  {Joushaghani}, \citenamefont {Iyer}, \citenamefont {Poon}, \citenamefont
  {Aitchison}, \citenamefont {De~Sterke}, \citenamefont {Wan},\ and\
  \citenamefont {Dignam}}]{joushaghani_generalized_2012_EDLExp}%
  \BibitemOpen
  \bibfield  {author} {\bibinfo {author} {\bibfnamefont {A.}~\bibnamefont
  {Joushaghani}}, \bibinfo {author} {\bibfnamefont {R.}~\bibnamefont {Iyer}},
  \bibinfo {author} {\bibfnamefont {J.~K.~S.}\ \bibnamefont {Poon}}, \bibinfo
  {author} {\bibfnamefont {J.~S.}\ \bibnamefont {Aitchison}}, \bibinfo {author}
  {\bibfnamefont {C.~M.}\ \bibnamefont {De~Sterke}}, \bibinfo {author}
  {\bibfnamefont {J.}~\bibnamefont {Wan}},\ and\ \bibinfo {author}
  {\bibfnamefont {M.~M.}\ \bibnamefont {Dignam}},\ }\href
  {https://doi.org/10.1103/PhysRevLett.109.103901} {\bibfield  {journal}
  {\bibinfo  {journal} {Physical Review Letters}\ }\textbf {\bibinfo {volume}
  {109}},\ \bibinfo {pages} {103901} (\bibinfo {year} {2012})}\BibitemShut
  {NoStop}%
\bibitem [{\citenamefont {An}\ \emph {et~al.}(2021)\citenamefont {An},
  \citenamefont {Padavić}, \citenamefont {Meier}, \citenamefont {Hegde},
  \citenamefont {Ganeshan}, \citenamefont {Pixley}, \citenamefont
  {Vishveshwara},\ and\ \citenamefont {Gadway}}]{an_interactions_2021_GAAH}%
  \BibitemOpen
  \bibfield  {author} {\bibinfo {author} {\bibfnamefont {F.~A.}\ \bibnamefont
  {An}}, \bibinfo {author} {\bibfnamefont {K.}~\bibnamefont {Padavić}},
  \bibinfo {author} {\bibfnamefont {E.~J.}\ \bibnamefont {Meier}}, \bibinfo
  {author} {\bibfnamefont {S.}~\bibnamefont {Hegde}}, \bibinfo {author}
  {\bibfnamefont {S.}~\bibnamefont {Ganeshan}}, \bibinfo {author}
  {\bibfnamefont {J.}~\bibnamefont {Pixley}}, \bibinfo {author} {\bibfnamefont
  {S.}~\bibnamefont {Vishveshwara}},\ and\ \bibinfo {author} {\bibfnamefont
  {B.}~\bibnamefont {Gadway}},\ }\href
  {https://doi.org/10.1103/PhysRevLett.126.040603} {\bibfield  {journal}
  {\bibinfo  {journal} {Physical Review Letters}\ }\textbf {\bibinfo {volume}
  {126}},\ \bibinfo {pages} {040603} (\bibinfo {year} {2021})}\BibitemShut
  {NoStop}%
\bibitem [{\citenamefont {Kraus}\ and\ \citenamefont
  {Zilberberg}(2012)}]{kraus_topological_2012-1_topology_IAAF}%
  \BibitemOpen
  \bibfield  {author} {\bibinfo {author} {\bibfnamefont {Y.~E.}\ \bibnamefont
  {Kraus}}\ and\ \bibinfo {author} {\bibfnamefont {O.}~\bibnamefont
  {Zilberberg}},\ }\href {https://doi.org/10.1103/PhysRevLett.109.116404}
  {\bibfield  {journal} {\bibinfo  {journal} {Physical Review Letters}\
  }\textbf {\bibinfo {volume} {109}},\ \bibinfo {pages} {116404} (\bibinfo
  {year} {2012})}\BibitemShut {NoStop}%
\bibitem [{\citenamefont {Goblot}\ \emph {et~al.}(2020)\citenamefont {Goblot},
  \citenamefont {Štrkalj}, \citenamefont {Pernet}, \citenamefont {Lado},
  \citenamefont {Dorow}, \citenamefont {Lemaître}, \citenamefont {Le~Gratiet},
  \citenamefont {Harouri}, \citenamefont {Sagnes}, \citenamefont {Ravets},
  \citenamefont {Amo}, \citenamefont {Bloch},\ and\ \citenamefont
  {Zilberberg}}]{goblot_emergence_2020_IAAF}%
  \BibitemOpen
  \bibfield  {author} {\bibinfo {author} {\bibfnamefont {V.}~\bibnamefont
  {Goblot}}, \bibinfo {author} {\bibfnamefont {A.}~\bibnamefont {Štrkalj}},
  \bibinfo {author} {\bibfnamefont {N.}~\bibnamefont {Pernet}}, \bibinfo
  {author} {\bibfnamefont {J.~L.}\ \bibnamefont {Lado}}, \bibinfo {author}
  {\bibfnamefont {C.}~\bibnamefont {Dorow}}, \bibinfo {author} {\bibfnamefont
  {A.}~\bibnamefont {Lemaître}}, \bibinfo {author} {\bibfnamefont
  {L.}~\bibnamefont {Le~Gratiet}}, \bibinfo {author} {\bibfnamefont
  {A.}~\bibnamefont {Harouri}}, \bibinfo {author} {\bibfnamefont
  {I.}~\bibnamefont {Sagnes}}, \bibinfo {author} {\bibfnamefont
  {S.}~\bibnamefont {Ravets}}, \bibinfo {author} {\bibfnamefont
  {A.}~\bibnamefont {Amo}}, \bibinfo {author} {\bibfnamefont {J.}~\bibnamefont
  {Bloch}},\ and\ \bibinfo {author} {\bibfnamefont {O.}~\bibnamefont
  {Zilberberg}},\ }\href {https://doi.org/10.1038/s41567-020-0908-7} {\bibfield
   {journal} {\bibinfo  {journal} {Nature Physics}\ }\textbf {\bibinfo {volume}
  {16}},\ \bibinfo {pages} {832} (\bibinfo {year} {2020})}\BibitemShut
  {NoStop}%
\bibitem [{\citenamefont {Verbin}\ \emph {et~al.}(2015)\citenamefont {Verbin},
  \citenamefont {Zilberberg}, \citenamefont {Lahini}, \citenamefont {Kraus},\
  and\ \citenamefont {Silberberg}}]{Verbin_PhysRevB.91.064201_IAAF_pumping}%
  \BibitemOpen
  \bibfield  {author} {\bibinfo {author} {\bibfnamefont {M.}~\bibnamefont
  {Verbin}}, \bibinfo {author} {\bibfnamefont {O.}~\bibnamefont {Zilberberg}},
  \bibinfo {author} {\bibfnamefont {Y.}~\bibnamefont {Lahini}}, \bibinfo
  {author} {\bibfnamefont {Y.~E.}\ \bibnamefont {Kraus}},\ and\ \bibinfo
  {author} {\bibfnamefont {Y.}~\bibnamefont {Silberberg}},\ }\href
  {https://doi.org/10.1103/PhysRevB.91.064201} {\bibfield  {journal} {\bibinfo
  {journal} {Phys. Rev. B}\ }\textbf {\bibinfo {volume} {91}},\ \bibinfo
  {pages} {064201} (\bibinfo {year} {2015})}\BibitemShut {NoStop}%
\end{thebibliography}%

\end{document}

%\clearpage
\onecolumngrid

\section*{Supplementary Information}
\setcounter{section}{0}
\setcounter{figure}{0}
\setcounter{section}{0}
\renewcommand{\theequation}{S\arabic{equation}}
\renewcommand{\thefigure}{S\arabic{figure}}
\makeatletter
\renewcommand{\bibnumfmt}[1]{[S#1]}
\renewcommand{\citenumfont}[1]{S#1}

\section{Numerical modeling of density evolution in the nonlinear Aubry-Andr\'e-Harper model}

In this section we present numerical simulations of the nonlinear (weakly interacting) Aubry-Andr\'e-Harper (AAH) model. These calculations provide evidence that the expansion speed of the second moment of a Gaussian initial state is approximately linearly dependent on the quasi-disorder strength $\lambda:= \Delta/J$, justifying the Bessel-like form of Fig.~2c in the main text. The dynamics can be described by the nonlinear Schr\"odinger equation
\[
i \partial_t \psi_j = -\left( \psi_{j+1} + \psi_{j-1} \right) + \lambda V_j(\varphi(t)) \psi_j + g\vert \psi_j \vert^2 \psi_j,
\]
where $g$ is the mean-field interaction strength in units of the tunneling strength $J$ and $V_j(\varphi(t)) = \cos(2\pi\alpha j + \varphi(t))$ is the AAH quasiperiodic potential. Here the time is units of the tunneling time $T_J := \hbar/J$.  

The numerical simulations use a second order split-step Fourier pseudospectral method with a system size $L = 2^{14}$ and a time step of $0.01 T_J$. The initial state is a Gaussian wave packet with an root-mean-squared (rms) width of $9 \, \mu m$. The appropriate value of $g$ depends on atomic properties and details of the experimental setup such as the strength of transverse confinement. The simulations presented here use $g = 95$, chosen close to the estimated experimental value. We note that this value is not in the self-trapping regime, which would require a much stronger interaction strength.

\begin{figure}[htb]
    \centering
    \includegraphics[width = \linewidth]{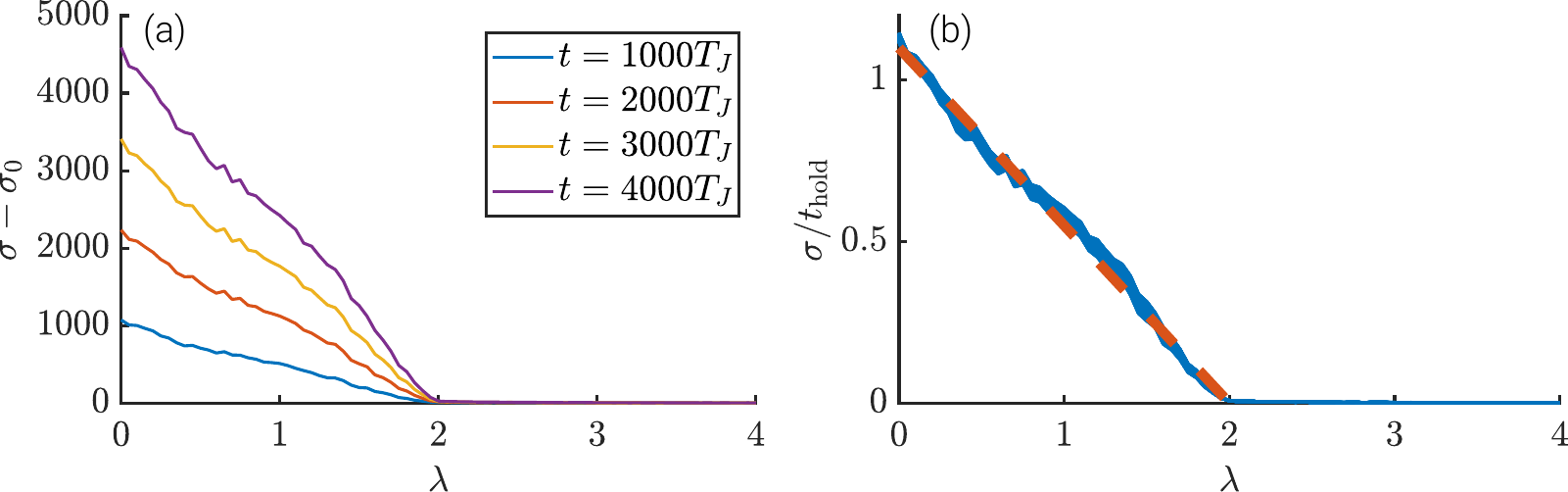}
    \caption{The rms width and expansion speed of an mean-field interacting AAH model averaged over $10$ different realizations. In \textbf{(a)}, we plot the late time rms width at different expansion times, all showing an approximately linear dependence on $\lambda$. In \textbf{(b)}, the blue curves are the computed average expansion speed at different hold times $t_\mathrm{hold}$ after $2000T_J$, which all collapse into almost a single curve. The red dashed line shows the linear dependence $0.55 \times (2 -\lambda)$. }
    \label{fig:apdx_expansion}
\end{figure}

The late-time simulation results are shown in Fig.~\ref{fig:apdx_expansion}. We computed the (rms) width $\sigma = \sqrt{m_2}$ of the wave function where $m_2 = \sum_j (j-j_0)^2\vert \psi_j\vert^2$ is the second moment. We observe that the final rms width depends roughly linearly on $\lambda$ in the delocalized regime. The approximate linear dependence can also be clearly observed in the ``expansion speed", defined as the ratio of the rms width $\sigma$ and the expansion time $t_\mathrm{hold}$. If this quantity is plotted versus $\lambda$, the late-time rms widths at different expansion times collapse into a single line, which is well approximated by the phenomenological relation $\sigma/t_\mathrm{hold} = 0.55 \times (2 -\lambda)$ in the delocalized regime. 

These results provide numerical evidence supporting the measurement of the Bessel function dependence using the late-time width of the condensate, as shown in figure 2c in the main text. We note that these simulations do not explain the deviations from the Bessel form visible at small phason amplitude in that figure, which may point to beyond-mean-field effects or experimental imperfections. 

 {
\section{Exact destruction of localization for generic bounded quasiperiodic systems}\label{a1}
}

As a demonstration of the utility of the superspace picture described in the main text, here we use it to predict and numerically verify the exact coherent phasonic destruction of localization for a generic bounded quasiperiodic potential in the high-driving-frequency regime. For simplicity we consider one-dimensional systems with nearest-neighbor hopping. The dynamics can be described by the discrete Schr\"odinger equation
\[
i \partial_t \psi_j = -\left( \psi_{j+1} + \psi_{j-1} \right) + \lambda V_j(\varphi(t)) \psi_j,
\]
where $\lambda$ is the strength of the quasiperiodic potential in units of the tunneling strength, $\psi_j$ is the probability amplitude of the wave function in the Wannier basis at site $j$, $V_j (\varphi(t))$ is the quasiperiodic potential, and time is measured in units of the tunneling time. We consider bounded quasiperiodic potentials that can be expanded into Fourier series \cite{ganeshan_nearest_2015_GAAH, devakul_anderson_2017_Superlattice}
\[
V_j = \sum_{n} a_n \cos n (2\pi\alpha j +\varphi(t)).
\]
In the superspace picture, each term in the expansion corresponds to a tunneling to the $n$-th neighboring site along the extra dimension. For example, the $n = 1$ term, which is just the AAH potential, corresponds to nearest-neighbor hopping; the $n = 2$ term corresponds to  next-nearest-neighbor hopping along the extra dimension, and so on. 

\begin{figure}[htb]
    \centering
    \includegraphics[width = \linewidth]{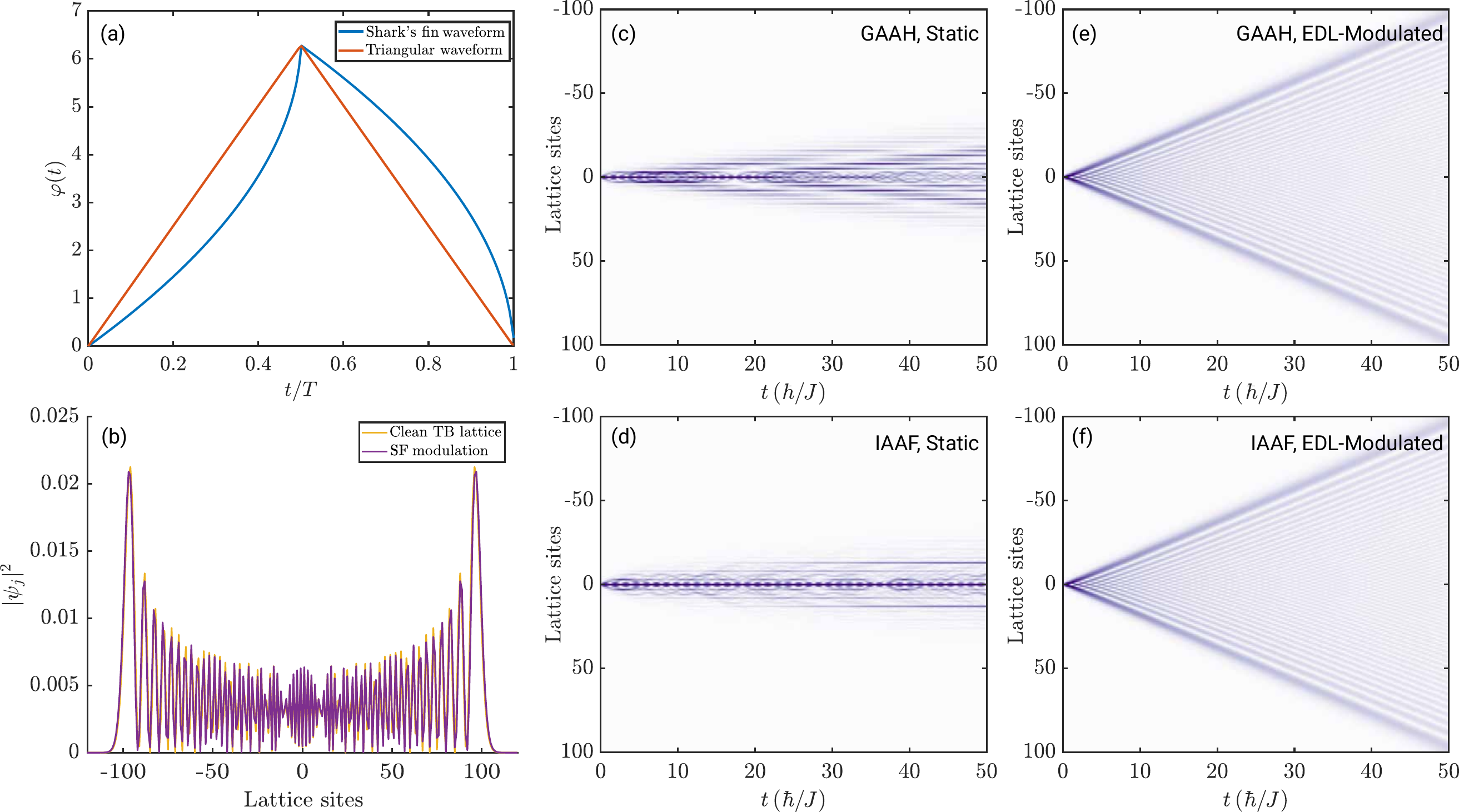}
    \caption{Numerical simulations showing the exact phasonic destruction of localization for two generic quasiperiodic models. (a) Comparison of the two modulation schemes $\varphi_\mathrm{tri}(t)$ and $\varphi_\mathrm{SF}(t)$. For the SF modulation we chose $b = -0.95$. (b) Comparison of the probability distributions $|\psi_j|^2$ of a clean nearest-neighbor tight-binding (TB) lattice (yellow) and of a quasiperiodic potential (here GAAH) under SF-phason modulation (purple) after $ t = 50 T_J$ expansion, showing excellent agreement. (c,d) Simulations of wave function expansion in static quasiperiodic potentials. (e,f) Simulations of wave function expansion in SF-phason modulated quasiperiodic potentials at $\varphi_0 = 2\pi$, displaying the destruction of localization in both models. For the GAAH model, we used $\lambda_\mathrm{GAAH} = -2$ and $a = -0.75$. At these parameters, the static GAAH model has a single-particle mobility edge. For the IAAF model, we used $\lambda_\mathrm{IAAF} = -1.8$ and $\beta_\mathrm{IAAF} = 10^5$ to reach the Fibonacci limit. The IAAF model is thus in the critical phase. All simulations started with a single-site excitation as the initial state and are driven with $\hbar \omega = 150J$. The incommensurate ratio $\alpha$ is set to the golden ratio $\alpha = (\sqrt{5}-1)/2$. The numerical integration is carried out with a second-order split-step method with the time step being $2.5\times 10^{-4} T_J$.}
    \label{fig:apdx_EDL}
\end{figure}

The phasonic destruction of localization in the 1D physical space corresponds to dynamic localization along the extra dimension in the superspace. The problem is thus transformed into finding the conditions of dynamical localization for a tight-binding lattice with arbitrary long-range hopping. This naturally connects to the concept of {\it exact dynamical localization} (EDL) \cite{dignam_conditions_2002_EDL, domachuk_dynamic_2002_EDL, eckardt_exploring_2009_DL, joushaghani_generalized_2012_EDLExp}. % developed more than a decade ago. 
Because a monochromatic sinusoidal modulation leads to dynamic localization only for the case of nearest-neighbor tunneling, correspondingly, a sinusoidal phasonic modulation leads to destruction of localization only for the case of a simple AAH model \cite{dignam_conditions_2002_EDL, domachuk_dynamic_2002_EDL}. However, there are many non-sinusoidal waveforms that are known to support EDL in more general circumstances~\cite{dignam_conditions_2002_EDL, domachuk_dynamic_2002_EDL, joushaghani_generalized_2012_EDLExp}. Taken together with the superspace picture, this suggests the concept of \emph{exact phasonic destruction of localization}. Here, we consider two such waveforms. The first is a triangle wave, corresponding to a square-wave electric field
\[
\varphi_\mathrm{tri}(t) = \begin{cases}
   2 \varphi_0 t/T, & 0 < t \leq T/2 \\
   -2 \varphi_0 (t/T-1), & T/2 < t \leq T,
\end{cases}
\]
The second waveform we consider is the ``shark's-fin'' (SF) form $\varphi_\mathrm{SF} = \varphi_0 f(t)$ where 
\[
f(t) = \begin{cases}
    \frac{1}{2b} \left({\sqrt{(b-1)^2 + 8bt/T} + b-1}\right), & 0<t\leq T/2 \\
    1- \frac{1}{2b} \left({\sqrt{b^2 - 6b +1 + 8bt/T} + b-1}\right), & T/2 < t \leq T,
\end{cases} \label{eq:EDL}
\]
where $|b|<1$ is the tuning parameter \cite{domachuk_dynamic_2002_EDL}. These two waveforms are plotted in Fig. \ref{fig:apdx_EDL}(a). For both waveforms, EDL is expected to occur for $\varphi_0 = 2n\pi$ where $n$ is an integer.

 {
We verify the prediction of exact phasonic destruction of localization with these two waveforms by numerical integration of discrete Schr\"odinger equations for two well-studied quasiperiodic systems beyond the simple AAH model. The first is the generalized AAH model (GAAH) \cite{ganeshan_nearest_2015_GAAH, an_interactions_2021_GAAH}: 
\[
V_j^\mathrm{GAAH} (\varphi(t)) = \frac{\cos \left(2 \pi \alpha j + \varphi(t)\right)}{ 1 - a \cos  \left(2 \pi \alpha j + \varphi(t)\right)},
\]
where the tuning parameter $a \in (-1,1)$ controls how much it deviates from the AAH model. The GAAH model is self-dual and hosts a mobility edge, and has been experimentally studied using momentum-space lattices \cite{an_interactions_2021_GAAH}. 
}

 {
The second example is the interpolated Aubry-Andr\'e-Fibonacci model (IAAF) that smoothly connects the AAH model and the Fibonacci model \cite{kraus_topological_2012-1_topology_IAAF, goblot_emergence_2020_IAAF, Verbin_PhysRevB.91.064201_IAAF_pumping}:
\[
V_j^\mathrm{IAAF} (\varphi(t)) = - \frac{\tanh \left\{\beta \left( \cos (2\pi \alpha j + \varphi(t)) - \cos(\pi \alpha)\right)\right\}}{\tanh \beta}.
\]
where the tuning parameter $\beta$ interpolates the two limits: $\beta \rightarrow 0$ gives the AAH potential (up to a global shift), and $\beta \rightarrow \infty$ gives the Fibonacci quasicrystal. 
}

 {
Figure \ref{fig:apdx_EDL}(c-f) shows the results of the proposed phasonic modulation for exact destruction of localization of these two quasiperiodic models using $\varphi_\text{SF}(t)$. We have also verified separately that triangle wave modulation $\varphi_\mathrm{tri} (t)$ leads to the same results. Here the initial state is a single-site excitation at the site $j_0=0$ for all cases. As demonstrated clearly in the Fig. \ref{fig:apdx_EDL}(b) and (e,f), the EDL modulation protocol leads to typical ballistic expansion from a single-site excitation in these two quasiperiodic models. The probability amplitudes at time $t$ in all the modulated cases match the analytical results of a clean tight-binding lattice, $|\psi_j|^2 = \left|\mathcal{J}_{j-j_0}(2t) \right|^2$ within numerical errors (Fig. \ref{fig:apdx_EDL}(b)). We emphasize that this result, attained naturally via the superspace picture, is nontrivial, since simple sinusoidal modulation is not capable of perfectly destroying localization in these more complex quasiperiodic models.

}

\bibliographystyle{apsrev4-2}
\bibliography{refs}